\documentclass{article}
\usepackage[dvips]{graphicx}

\input{tcilatex}
\begin{document}

\title{Anisotropic Pressures at Ultra-stiff Singularities\\
and the Stability of Cyclic Universes}
\author{John D Barrow and Kei Yamamoto \\
%EndAName
DAMTP, Centre for Mathematical Sciences,\\
University of Cambridge,\\
Wilberforce Rd., Cambridge CB3 0WA\\
United Kingdom}
\maketitle

\begin{abstract}
We show that the inclusion of simple anisotropic pressures stops the
isotropic Friedmann universe being a stable attractor as an initial or final
singularity is approached when pressures can exceed the energy density. This
shows that the situation with isotropic pressures, studied earlier in the
context of cyclic and ekpyrotic cosmologies, is not generic, and Kasner-like
behaviour occurs when simple pressure anisotropies are present. We find all
the asymptotic behaviours and determine the dynamics when the anisotropic
principal pressures are proportional to the density. We expect distortions
and anisotropies to be significantly amplified through a simple cosmological
bounce in cyclic or ekpyrotic cosmologies when ultra-stiff pressures are
present.

PACs nos 98.80.Jk, 04.40.Nr, 04.50.+h, 11.25.Mj
\end{abstract}

\section{Introduction}

There has been strong interest in the cosmological consequences of admitting
a 'ultra-stiff' fluid, whose isotropic pressure exceeds its energy density
in the early stages of the universe. This situation could occur in a number
of scenarios created by attempts to develop non-singular descriptions of
spacetime which are applicable at arbitrarily early cosmological epochs. The
simplest example is that provided by a scalar field, $\phi $, with a
negative potential energy, $V(\phi )<0$, in a homogeneous and isotropic
universe, so that the pressure-density ratio is

\[
\frac{p}{\rho }\equiv \gamma -1=\frac{\frac{1}{2}\dot{\phi}^{2}-V}{\frac{1}{2%
}\dot{\phi}^{2}+V}, 
\]%
and for $V<0$ we can have $\ p/\rho >1$ in a regime where $\left\vert
V\right\vert >\frac{1}{2}\dot{\phi}^{2}$. This scenario can arise in the so
called ekpyrotic \cite{ek}, or cyclic, universe scenarios described by
Gasperini and Veneziano \cite{gasp} and Erickson et al \cite{erik}, but can
also occur in other theories containing effective scalar fields, for example
in some compactifications in higher-dimensional theories \cite{hull, witten}%
. As a result, several authors have investigated the cosmological
consequences of the presence of a simple, ultra-stiff, $\gamma >2$ perfect
fluid. Erickson et al \cite{erik} and Lidsey \cite{lid} showed that an
ultra-stiff perfect fluid renders the isotropic Friedmann universe stable as
an initial or final singularity is approached in general relativistic
cosmologies. This is quite different to the situation when $\gamma \leq 2$,
where isotropic expansion is unstable on approach to initial and (if
present) final singularities. Theorems were proved to establish this
stability, a form of cosmic no-hair theorem, for homogeneous and anisotropic
cosmologies, where the problem reduces to the analysis of ordinary
differential equations \cite{wain}. The results are intuitively obvious for
perfect fluids: there is no form of curvature or expansion anisotropy can
diverge faster than $a^{-6}$ as the mean scale factor $a(t)\rightarrow 0$,
but a fluid with $p=(\gamma -1)\rho >\rho $ the density will diverge
isotropically as $\rho \propto a^{-3\gamma }$ in the same limit and dominate
the anisotropic stresses in the $a\rightarrow 0$ limit when $\gamma >2$. If
a real fluid with rotational motion is present then the vortical energy
density, $\tilde{\Omega}^{2}$, goes to zero in the same limit, since by
conservation of angular momentum \cite{vort} it evolves as $\tilde{\Omega}%
^{2}\propto a^{2(3\gamma -4)}\rightarrow 0$ as $a\rightarrow 0$ for $\gamma
>2$ and $\tilde{\Omega}^{2}/\rho \propto a^{9\gamma -10}\rightarrow 0$ for $%
\gamma >10/9$.

This stability of the isotropy and homogeneity of the Friedmann solution as $%
a\rightarrow 0$ is a very important ingredient for cyclic cosmologies
because it permits a regular transition from collapse to expansion from
cycle to cycle. If the isotropic expansion were unstable as $a\rightarrow 0$
then huge irregularities and anisotropies would accumulate and the
successive cycles would be very different and increasingly anisotropic. In
particular, attempts to follow small inhomogeneous perturbations through the
bounce would fail \cite{perts}.

In this paper we will show that the studies of the likely evolution of
cosmological models as $a\rightarrow 0$ when pressures can exceed energy
density have only considered a restricted situation in which the ultra-stiff
pressures are isotropic. When anisotropic pressures are also present the
stability of the Friedmann solution fails as the initial and final
singularity is approached and the cyclic scenario loses this appealing
feature.

\section{Anisotropic pressures}

Earlier analyses of the $p>\rho $ cosmological problem have restricted their
attention to the situation where the pressure is isotropic \cite{erik,lid}.
However, if anisotropic pressures are present we should also expect the
principal pressure components $(p_{1},p_{2},p_{3})\ $ to be able to range
over values that exceed $\rho $. This corresponds to a violation of the
dominant energy condition, (which requires $T^{00}\geq \left\vert
T^{ab}\right\vert $ for each $a,b$, \cite{he}). Physically, anisotropic
pressures are to accompany anisotropic expansion at very high energies
because asymptotically-free interactions become collisionless when $%
T>10^{15}GeV$ and gravitons will be collisionless below the Planck energy
scale ($T<10^{19}GeV$). Thus, if there is any expansion anisotropy, these
collisionless particles will redshift (or even blueshift) at different rates
in different directions and create significant anisotropic stresses with
some (or all) experiencing $p_{i}>\rho $. The evolution as the universe
expands is complicated and has been studied in detail for the $p<\rho $
situation, especially for the radiation and dust-dominated eras when $p/\rho
=1/3$ or $1$, for evolution away from the singularity to the present \cite%
{skew, lukash}. Here, we will be interested in the behaviour of the
cosmology in the limit of approach to the initial singularity with isotropic
and anisotropic fluids both present. In order to establish a failure of the
cosmic no-hair approach to the Friedmann solution it suffices to consider
only the simple Bianchi type I universe. More complicated Bianchi type
universes will provide further scope for anisotropic stresses to dominate
but we will not include them here.

\section{Bianchi I with ultra-stiff perfect fluid}

Throughout this paper, we make use of orthonormal frame formalism developed
by Ellis et al \cite{ellis}. As a simple introduction we will consider an
anisotropic Bianchi type I universe containing a perfect fluid with
ultra-stiff equation of state 
\begin{equation}
T_{ab}^{I}=\rho \left\{ u_{a}u_{b}+(\gamma -1)g_{ab}\right\} ,
\end{equation}%
where $\gamma >2$ is a constant, as above. We assume the fluid 4-velocity
vector $\mathbf{u}$ is orthogonal to the homogeneous space-like
hypersurfaces. We use an orthonormal tetrad so that Einstein equations and
Bianchi identities are put into the following form: 
\begin{eqnarray*}
\dot{H} &=&-H^{2}-\frac{2}{3}\sigma ^{2}-\frac{1}{6}(3\gamma -2)\rho , \\
\dot{\sigma}_{\alpha \beta } &=&-3H\sigma _{\alpha \beta }+2\epsilon _{\ \
(\alpha }^{\mu \nu }\sigma _{\beta )\mu }\Omega _{\nu }, \\
\rho &=&3H^{2}-\sigma ^{2}, \\
\dot{\rho} &=&-3\gamma H\rho ,
\end{eqnarray*}%
where 
\begin{equation}
\sigma ^{2}\equiv \frac{1}{2}\sigma _{\alpha \beta }\sigma ^{\beta \alpha }.
\end{equation}%
Here, $H$ is Hubble parameter (mean expansion rate), $\sigma _{\alpha \beta
} $ is the traceless shear expansion tensor, and $\Omega _{\alpha }$ is
angular velocity of the tetrad frame with respect to a Fermi-propagated
frame. We define proper time $t$ by 
\[
\frac{\partial }{\partial t}\equiv \mathbf{u} 
\]%
and denote the time derivative with respect to $t$ by overdot. Greek indices
are used for space components of the tetrad. We can use the freedom to
choose the space tetrad to diagonalise $\sigma _{\alpha \beta }$. The shear
evolution equations then imply 
\[
\Omega _{\alpha }=0. 
\]

As is the case for the FLRW universe, these equations are not independent.
To single out the dynamical degrees of freedom, we introduce
expansion-normalised variables: 
\begin{eqnarray*}
\Sigma _{\alpha \beta } &\equiv &\frac{\sigma _{\alpha \beta }}{H}, \\
\Omega &\equiv &\frac{\rho }{3H^{2}}, \\
q &\equiv &-1-\frac{\dot{H}}{H^{2}}.
\end{eqnarray*}%
Since we are interested in the evolution of the universe during its
expansion, we also define a dimensionless time $\tau $ by 
\[
\frac{d\tau }{dt}\equiv H, 
\]%
and a mean scale factor by 
\[
\frac{\dot{l}}{l}\equiv H. 
\]%
We see that 
\[
l\propto e^{\tau } 
\]%
and therefore the initial singularity ($l=0$) corresponds to $\tau
\rightarrow -\infty $. Denoting $\tau $ derivatives by primes, we have 
\begin{eqnarray*}
\Sigma _{\alpha \beta }^{\prime } &=&(q-2)\Sigma _{\alpha \beta }, \\
\Omega &=&1-\Sigma ^{2}\equiv 1-\frac{1}{2}\Sigma _{\alpha \beta }\Sigma
^{\beta \alpha }, \\
q &=&2\Sigma ^{2}+\frac{1}{2}(3\gamma -2)\Omega ,
\end{eqnarray*}%
with an auxiliary equation 
\[
\Omega ^{\prime }=(2q-3\gamma +2)\Omega , 
\]%
and a decoupled equation 
\[
H^{\prime }=-(1+q)H. 
\]%
These equations can be readily integrated to give 
\begin{eqnarray*}
\Omega &=&\frac{1}{2}+\frac{1}{2}\tanh \left( -\frac{3}{2}(\gamma -2)\tau
\right) , \\
\Sigma _{\alpha \beta } &=&\frac{\Sigma _{0\alpha \beta }}{\sqrt{2}}e^{\frac{%
3}{4}(\gamma -2)\tau }\cosh ^{-\frac{1}{2}}\left( -\frac{3}{2}(\gamma
-2)\tau \right) , \\
H &=&H_{0}e^{-\frac{3}{2}(\gamma +2)\tau }\cosh ^{\frac{1}{2}}\left( -\frac{3%
}{2}(\gamma -2)\tau \right) ,
\end{eqnarray*}%
where $\Sigma _{0\alpha \beta }$ are integration constants satisfying 
\begin{eqnarray*}
\Sigma _{0\ \alpha }^{\ \alpha } &=&0, \\
\frac{1}{2}\Sigma _{0\alpha \beta }\Sigma _{0}^{\beta \alpha } &=&1,
\end{eqnarray*}%
while the constant $H_{0}$ can be absorbed into the definition of time.
Since $\gamma >2$, we see that $\Sigma <<\Omega $ when the universe is
"small", near any initial singularity at $l=0$.

If we take the general Bianchi I metric 
\begin{equation}
ds^{2}=-dt^{2}+a(t)^{2}dx^{2}+b(t)^{2}dy^{2}+c(t)^{2}dz^{2},
\label{eq:metric}
\end{equation}%
the corresponding kinematic variables are given by 
\begin{eqnarray*}
H &=&\frac{1}{3}\left( \frac{\dot{a}}{a}+\frac{\dot{b}}{b}+\frac{\dot{c}}{c}%
\right) , \\
\sigma _{11} &=&\frac{\dot{a}}{a}-H, \\
\sigma _{22} &=&\frac{\dot{b}}{b}-H, \\
\sigma _{33} &=&\frac{\dot{c}}{c}-H, \\
\sigma _{\alpha \beta } &=&0\ \ \ \ \ \ (\alpha \neq \beta ), \\
\Omega _{\alpha } &=&0.
\end{eqnarray*}%
The results for expansion normalised variables are translated into metric
language as 
\begin{eqnarray*}
a &\propto &e^{(1-\Sigma _{011})\tau }\left( 1+\sqrt{1+e^{-3(\gamma -2)\tau }%
}\right) ^{-\frac{2\Sigma _{011}}{3(\gamma -2)}} \\
b &\propto &e^{(1-\Sigma _{022})\tau }\left( 1+\sqrt{1+e^{-3(\gamma -2)\tau }%
}\right) ^{-\frac{2\Sigma _{022}}{3(\gamma -2)}} \\
c &\propto &e^{(1-\Sigma _{033})\tau }\left( 1+\sqrt{1+e^{-3(\gamma -2)\tau }%
}\right) ^{-\frac{2\Sigma _{033}}{3(\gamma -2)}}.
\end{eqnarray*}%
When $\tau $ $\rightarrow $ $-\infty $, the asymptotic behaviour is 
\[
a,b,c\sim e^{\tau }\propto l\rightarrow 0 
\]%
and the universe becomes isotropic. On the other hand, when $\tau $ is
large, we have 
\[
(a,b.c)\sim (e^{(1-\Sigma _{011})\tau },e^{(1-\Sigma _{022})\tau
},e^{(1-\Sigma _{033})\tau })\propto (l^{1-\Sigma _{011}},\ l^{1-\Sigma
_{022}},l^{1-\Sigma _{033}})\mathrm{.} 
\]%
This is the Kasner vacuum metric with Kasner indices 
\[
p_{i}=\frac{1}{3}(1-\Sigma _{0ii}). 
\]%
We conclude that the universe become fluid-dominated and isotropic on
approach to the singularity. This is in accord with the results of Erickson
et al \cite{erik} and Lidsey \cite{lid}.

\section{Anisotropic ultra-stiff Bianchi I universes}

We will now investigate how the above result is modified by the presence of
an ultra-stiff anisotropic fluid in addition to an ultra-stiff isotropic
fluid. As a simple model we take the total energy-momentum tensor to be 
\begin{eqnarray}
T_{ab} &=&T_{ab}^{I}+T_{ab}^{A}, \\
T_{ab}^{A} &=&\mu \{u_{a}u_{b}+(\gamma _{\ast }-1)g_{ab}+\mathcal{P}_{ab}\},
\end{eqnarray}%
where $\gamma _{\ast }>2$ and $\mathcal{P}_{ab}$ is a constant traceless
(and has to be symmetric to be consistent with the metric) tensor describing
the ratio between energy density and anisotropic stresses (note that
momentum flow is not allowed in Bianchi I). This form for the anisotropic
stress tensor is characteristic of electromagnetic fields and many other
anisotropic stresses in the early universe (see ref \cite{skew,bmaart} for a
fuller discussion) although cosmological Yang-Mills fields require a
different form of anisotropic stress tensor \cite{ym}. The governing
equations are modified and become 
\begin{eqnarray}
\dot{H} &=&-H^{2}-\frac{2}{3}\sigma ^{2}-\frac{1}{6}(3\gamma -2)\rho -\frac{1%
}{6}(3\gamma _{\ast }-2)\mu , \\
\dot{\sigma}_{\alpha \beta } &=&-3H\sigma _{\alpha \beta }+2\epsilon _{\ \
(\alpha }^{\mu \nu }\sigma _{\beta )\mu }\Omega _{\nu }+\mu \mathcal{P}%
_{\alpha \beta },  \label{eq:shear} \\
\rho +\mu &=&3H^{2}-\sigma ^{2}, \\
\dot{\rho} &=&-3\gamma H\rho , \\
\dot{\mu} &=&-3\gamma _{\ast }H\mu -\sigma _{\alpha \beta }\mathcal{P}%
^{\beta \alpha }\mu ,
\end{eqnarray}%
Again, diagonalising shear tensor $\sigma _{\alpha \beta }$ gives 
\[
0=(\sigma _{33}-\sigma _{22})\Omega _{1}+\mu \mathcal{P}_{23}\ \ \ \mathrm{%
etc,} 
\]%
as the off-diagonal parts of (\ref{eq:shear}) are just defining relations
for $\Omega _{\alpha }$ and are non-dynamical. For later convenience, we
introduce 
\begin{eqnarray}
\sigma _{+} &\equiv &\frac{1}{2}(\sigma _{22}+\sigma _{33}), \\
\sigma _{-} &\equiv &\frac{1}{2\sqrt{3}}(\sigma _{22}-\sigma _{33}),
\end{eqnarray}%
where the pre-factors are chosen so that 
\[
\sigma ^{2}=3(\sigma _{+}^{2}+\sigma _{-}^{2}). 
\]%
These variables are sufficient to describe shear expansion tensor because
they are traceless. The sub-case with $\sigma _{-}=0$ is the axisymmetric
Bianchi type I metric. The remaining equations are then rewritten as 
\begin{eqnarray}
\dot{H} &=&-H^{2}-2(\sigma _{+}^{2}+\sigma _{-}^{2})-\frac{1}{6}(3\gamma
-2)\rho -\frac{1}{6}(3\gamma _{\ast }-2)\mu , \\
\dot{\sigma}_{\pm } &=&-3H\sigma _{\pm }+\mathcal{P}_{\pm }\mu , \\
\dot{\mu} &=&-3\gamma _{\ast }H\mu -6(\mathcal{P}_{+}\sigma _{+}+\mathcal{P}%
_{-}\sigma _{-})\mu , \\
\dot{\rho} &=&-3\gamma H\rho , \\
3H^{2} &=&\rho +\mu +3(\sigma _{+}^{2}+\sigma _{-}^{2}),
\end{eqnarray}%
with the definitions 
\begin{eqnarray}
\mathcal{P}_{+} &\equiv &\frac{1}{2}(\mathcal{P}_{22}+\mathcal{P}_{33}), \\
\mathcal{P}_{-} &\equiv &\frac{1}{2\sqrt{3}}(\mathcal{P}_{22}-\mathcal{P}%
_{33}).
\end{eqnarray}%
describing the pressure anisotropies. As the previous section, we introduce
an expansion-normalised variable for the density: 
\[
Z\equiv \frac{\mu }{3H^{2}}. 
\]%
The equations are cast into the set 
\begin{eqnarray}
\Sigma _{\pm }^{\prime } &=&(q-2)\Sigma _{\pm }+3\mathcal{P}_{\pm }Z,
\label{eq:a} \\
\Omega ^{\prime } &=&(2q-3\gamma +2)\Omega ,  \label{eq:b} \\
Z^{\prime } &=&(2q-3\gamma _{\ast }+2-6\mathcal{P}_{+}\Sigma _{+}-6\mathcal{P%
}_{-}\Sigma _{-})Z,  \label{eq:c} \\
q &=&2(\Sigma _{+}^{2}+\Sigma _{-}^{2})+\frac{1}{2}(3\gamma -2)\Omega +\frac{%
1}{2}(3\gamma _{\ast }-2)Z,  \label{eq:d} \\
1 &=&\Omega +Z+\Sigma _{+}^{2}+\Sigma _{-}^{2}.  \label{eq:e}
\end{eqnarray}%
These equations no longer solve exactly but the qualitative and asymptotic
behaviour of the system can be determined.

\section{The dynamical system}

The system described by equations (\ref{eq:a}) to (\ref{eq:e}) is
three-dimensional and compact because of the constraint (\ref{eq:e}). We
denote the entire system by $\overline{B_{A}(\mathrm{I})}$. First, we have
to find lower-dimensional invariant subsets. It turns out there are three
which play important roles. They are defined as follows:

\begin{description}
\item[Bianchi I] 
\[
B(\mathrm{I})\equiv \{Z=0\} 
\]

\item[Anisotropic fluid ] 
\[
A(\mathrm{I})\equiv \{\Omega =0\} 
\]

\item[Aligned to stress 'vector' $(\mathrm{\mathcal{P}_{+},\mathcal{P}_{-})}$%
] 
\[
\Pi (\mathrm{{I})\equiv \{\Sigma _{+}\mathcal{P}_{-}=\mathcal{P}_{+}\Sigma
_{-}\}.} 
\]
\end{description}

All of these are two-dimensional subsets; $B(\mathrm{{I})}$ and $A(\mathrm{{I%
})}$ lie in the boundary of the whole space. If we take $Z$ and $\Sigma
_{\pm }$ as independent variables, $\overline{B_{A}(\mathrm{{I})}}$ looks
like an inverted bowl with parabolic surface and $\Pi (\mathrm{{I})}$ is a
vertical slice passing through the axis of symmetry. The subset $B(\mathrm{{I%
})}$ is the base of the bowl and was already discussed in section 3.

At this point, it is instructive to introduce polar coordinate for shear
variables defined by 
\begin{eqnarray*}
\Sigma _{+} &\equiv &\Sigma \cos \phi , \\
\Sigma _{-} &\equiv &\Sigma \sin \phi .
\end{eqnarray*}%
Accordingly, we parametrise the shear tensor by 
\begin{eqnarray*}
\mathcal{P}_{+} &\equiv &-\frac{\gamma _{\ast }-2}{2}r\cos \theta , \\
\mathcal{P}_{-} &\equiv &-\frac{\gamma _{\ast }-2}{2}r\sin \theta .
\end{eqnarray*}%
In these coordinates $\phi $ measures the angle around the Kasner circle and 
$\phi =\theta $ (and $\phi =\theta +\pi $) corresponds to $\Pi (I)$. The
governing equations are then further simplified to 
\begin{eqnarray}
\Sigma ^{\prime } &=&(q-2)\Sigma -\frac{3}{2}(\gamma _{\ast }-2)rZ\cos (\phi
-\theta ), \\
\phi ^{\prime } &=&\frac{3Z}{2\Sigma }(\gamma _{\ast }-2)r\sin (\phi -\theta
), \\
\Omega ^{\prime } &=&(2q-3\gamma +2)\Omega , \\
Z^{\prime } &=&\left[ 2q-3\gamma _{\ast }+2+3(\gamma _{\ast }-2)r\Sigma \cos
(\phi -\theta )\right] Z, \\
q &=&2\Sigma ^{2}+\frac{1}{2}(3\gamma -2)\Omega +\frac{1}{2}(3\gamma _{\ast
}-2)Z, \\
1 &=&\Omega +Z+\Sigma ^{2}.
\end{eqnarray}

The second step is to find the equilibrium points of the system and
determine their stability properties. Time derivatives of all the normalised
variables vanish at an equilibrium point and it corresponds to a
self-similar solution. They are most conveniently characterised by defining
the parameter 
\begin{equation}
\alpha \equiv \frac{\gamma _{\ast }-\gamma }{\gamma _{\ast }-2},
\label{alpha}
\end{equation}%
which measures the stiffness of the anisotropic fluid compared to the
isotropic one.

We can now list all the equilibrium points, their metric interpretation, and
their eigenvalues:

\begin{description}
\item[FL equilibrium point ] defined by: 
\[
\Omega =1,\ \ \ Z=\Sigma =0 
\]%
Eigenvalues 
\[
\lambda _{1}=-3(\gamma _{\ast }-2)\alpha ,\ \ \ \lambda _{2}=\lambda _{3}=%
\frac{3}{2}(\gamma _{\ast }-2)(1-\alpha ) 
\]%
We denote this by $\mathcal{F}$.

\item[Kasner equilibrium points] defined by: 
\[
\Sigma =1,\ \ \ \Omega =Z=0,\ \ \ \phi =\mathrm{arbitrary\ constant} 
\]%
Eigenvalues 
\[
\lambda _{1}=0,\ \ \ \lambda _{2}=-3(\gamma _{\ast }-2)(1-\alpha ),\ \ \
\lambda _{3}=-3(\gamma _{\ast }-2)\left( 1-r\cos (\phi -\theta )\right) 
\]%
This set of equilibrium points, called the Kasner circle, will be denoted by 
$\mathcal{K}$.

\item[Anisotropic 1-fluid equilibrium point] defined by: 
\[
\Sigma =r,\ \ \ \phi =\theta ,\ \ \ \Omega =0,\ \ \ Z=1-r^{2} 
\]%
Eigenvalues 
\begin{eqnarray*}
\lambda _{1} &=&3(\gamma _{\ast }-2)(\alpha -r^{2}) \\
\lambda _{2} &=&\lambda _{3}=\frac{3}{2}(\gamma _{\ast }-2)(1-r^{2})
\end{eqnarray*}%
We denote this by $\mathcal{A}_{1}$. It becomes unphysical unless $0\leq
r\leq 1$ because of the constraint (\ref{eq:e}) and the need for positivity
of $\Sigma $ and $Z$. For $r=0$ it is identical to $\mathcal{F}$ and merges
into the point on the Kasner circle with $\phi =\theta $ when $r=1$. To work
out the scale factors in the same fashion as we did in section 3, we have to
restrict ourselves to the case $\mathcal{P}_{\alpha \beta }=0,(\alpha \neq
\beta )$ because the metric (\ref{eq:metric}) does not admit rotation or
off-diagonal shear. The scale factors are 
\[
a\propto t^{p_{1}},\ \ b\propto t^{p_{2}},\ \ c\propto t^{p_{3}} 
\]%
with 
\begin{eqnarray*}
p_{i} &=&\frac{1}{1+q}\left( 1-\frac{2\mathcal{P}_{ii}}{\gamma _{\ast }-2}%
\right) \\
1+q &=&\frac{3\gamma _{\ast }}{2}-\frac{1}{\gamma _{\ast }-2}(\mathcal{P}%
_{11}^{2}+\mathcal{P}_{22}^{2}+\mathcal{P}_{33}^{2})\ =\ \frac{3}{2}\gamma
_{\ast }(1-r^{2})+3r^{2}.
\end{eqnarray*}%
This has the same form as the Kasner solutions but instead of the usual
summation relation, the exponents satisfy 
\begin{eqnarray}
p_{1}+p_{2}+p_{3} &=&\frac{3}{1+q}  \label{sum1} \\
p_{1}^{2}+p_{2}^{2}+p_{3}^{2} &=&\frac{3}{(1+q)^{2}}(1+2r^{2}).
\end{eqnarray}

\item[Anisotropic 2-fluid equilibrium point] defined by: 
\[
\Sigma =\frac{\alpha }{r},\ \ \ \phi =\theta ,\ \ \ \Omega =1-\frac{\alpha }{%
r^{2}},\ \ \ Z=\frac{\alpha (1-\alpha )}{r^{2}} 
\]%
Eigenvalues 
\begin{eqnarray*}
\lambda _{1} &=&\frac{3}{4}(\gamma _{\ast }-2)\left[ 1-\alpha +\sqrt{%
(1-\alpha )\left( 1-9\alpha +8\frac{\alpha ^{2}}{r^{2}}\right) }\right] \\
\lambda _{2} &=&\frac{3}{4}(\gamma _{\ast }-2)\left[ 1-\alpha -\sqrt{%
(1-\alpha )\left( 1-9\alpha +8\frac{\alpha ^{2}}{r^{2}}\right) }\right] \\
\lambda _{3} &=&\frac{3}{2}(\gamma _{\ast }-2)(1-\alpha )
\end{eqnarray*}%
We denote this critical point by $\mathcal{A}_{2}$. For it to be physical we
require 
\begin{equation}
0\leq \alpha \leq r^{2}\leq 1  \label{ineq1}
\end{equation}%
or 
\begin{equation}
0\leq \alpha \leq 1\leq r.  \label{ineq2}
\end{equation}%
The metric is the same form as $\mathcal{A}_{1}$ with the Kasner exponents
given by 
\begin{eqnarray}
p_{1} &=&\frac{2}{3\gamma }\left[ 1-\alpha \frac{2\mathcal{P}_{11}}{\gamma
_{\ast }-2}\right]  \nonumber \\
p_{2} &=&\frac{2}{3\gamma }\left[ 1-\alpha \frac{2\mathcal{P}_{22}}{\gamma
_{\ast }-2}\right]  \nonumber \\
p_{3} &=&\frac{2}{3\gamma }\left[ 1-\alpha \frac{2\mathcal{P}_{33}}{\gamma
_{\ast }-2}\right]  \nonumber \\
p_{1}+p_{2}+p_{3} &=&\frac{2}{\gamma }  \label{sum2} \\
p_{1}^{2}+p_{2}^{2}+p_{3}^{2} &=&\frac{4}{3\gamma ^{2}}(1+2\alpha ^{2}r^{2}).
\end{eqnarray}
\end{description}

All of these points except for the Kasner circle lie in $\Pi (\mathrm{{I})}$.

\subsection{Summary of stability properties}

We summarise the stability of those equilibrium points for the case $\gamma
_{\ast }>2$ (ie the anisotropic fluid is ultra-stiff). This was the
situation in which the isotropic Friedmann-Lema\^{\i}tre model was shown to
be the single attractor on approach to the initial singularity in past
studies \cite{erik},\cite{lid}. The particular cases in which at least one
of the inequalities in (\ref{ineq1}) or (\ref{ineq2}) on $\alpha ,r$ or $%
r^{2}$ become equalities will be considered later. Note that our time
derivative, $\tau ,$ is defined so that the universe is getting larger
towards the future. Therefore, in the context of contracting universe, we
are interested in the past asymptotic behaviour with respect to $\tau $. In
terms of stability, the unstable equilibrium points are important.

\subsubsection{ $\protect\alpha <0 $, $r>1$}

In this case both $\mathcal{A}_1$ and $\mathcal{A}_2$ are located outside
physical domain. $\mathcal{F}$ is unstable and $\mathcal{K}$ has stable arc
and an arc of saddle points.

\subsubsection{ $\protect\alpha <0 $, $r<1$}

$A_2$ is not physical. $\mathcal{F}$ is unstable and $\mathcal{K}$ is
stable. $A_1$ is unstable in the invariant set $\Omega =0$ but a saddle
point in the interior of the entire space.

\subsubsection{ $0<\protect\alpha <1 $, $\protect\alpha < r^2 <1$}

All equilibrium points lie in the physical domain. $\mathcal{F}$ is unstable
in the invariant set Bianchi I and a saddle point in general. $\mathcal{A}_1$
is another saddle point. $\mathcal{K}$ is stable. $\mathcal{A}_2$ turns out
to be unstable and will be identified as the past attractor of the entire
system.

\subsubsection{ $0<\protect\alpha <1 $, $r^2<\protect\alpha $}

$\mathcal{A}_2$ is outside physical domain. Instead $A_1 $ becomes unstable
and replaces the role of $\mathcal{A}_2$ in the previous case.

\subsubsection{ $0<\protect\alpha <1 $, $r >1$}

In this case $\mathcal{A}_1$ is outside the domain while $\mathcal{A}_2$ is
physical and unstable. The difference is a part of Kasner circle becomes
saddle.

\subsubsection{ $\protect\alpha >1$, $r<1$}

The isotropic fluid is no longer stiff and $\mathcal{A}_{2}$ becomes
non-physical. $\mathcal{F}$ is the global future attractor, $\mathcal{K}$ is
saddle and $\mathcal{A}_{1}$ is the past attractor.

\subsubsection{ $\protect\alpha >1$, $r >1$}

The situation is the same as the previous one except for non-physical $%
\mathcal{A}_1$. A part of Kasner circle becomes the past attractor.

\subsection{Monotone functions}

Finally, we list some monotone functions crucial to understand the
asymptotic behaviour.

\begin{description}
\item[$\protect\phi$ ] Monotone for $r\neq 0$ in $\overline{B_A (\mathrm{{I})%
}} / \left( \Pi (\mathrm{{I}) \cup B({I}) \cup \{ \Sigma =0 \} }\right)$. 
\newline
Increasing for $0 < \phi - \theta < \pi$ and decreasing for $-\pi < \phi
-\theta <0$.

\item[$\Omega $] Monotone decreasing for $\alpha <0$ in $\overline{B_{A}(%
\mathrm{{I})}}/A(\mathrm{{I})}$. \newline
For $\alpha =0$, it is semi-monotone decreasing and $\Omega ^{\prime }=0$
iff $\Sigma =0$.
\end{description}

From the monotonicity of $\phi $ we conclude that it is sufficient to look
at $\Pi (\mathrm{{I})}$ and $B(\mathrm{{I})}$ in order to determine the
asymptotic behaviour . In particular, for $\alpha <1,r<1$, the past
attractor (if it exists) must lie in $\Pi (\mathrm{{I})}$ because the Kasner
circle is stable.

\subsection{Invariant set $A(\mathrm{{I})}$}

For completeness and to get a flavour of the dynamics involving anisotropic
fluid, we consider the invariant set $A(\mathrm{{I})}$. Setting $\Omega =0$,
the equations read 
\begin{eqnarray}
\Sigma _{\pm }^{\prime } &=&(q-2)\Sigma _{\pm }+3\mathcal{P}_{\pm }Z, \\
q &=&2(\Sigma _{+}^{2}+\Sigma _{-}^{2})+\frac{1}{2}(3\gamma _{\ast }-2)Z,
\label{eq:q} \\
Z &=&1-(\Sigma _{+}^{2}+\Sigma _{-}^{2}).  \label{eq:omega}
\end{eqnarray}%
We can easily eliminate $q$ and $Z$ to obtain 
\begin{equation}
\Sigma _{\pm }^{\prime }=-3(1-\Sigma _{+}^{2}-\Sigma _{-}^{2})\left[ \frac{1%
}{2}(2-\gamma _{\ast })\Sigma _{\pm }-\mathcal{P}_{\pm }\right] .
\label{eq:A}
\end{equation}%
To see the asymptotic behavior, we recast (\ref{eq:A}) into 
\[
\left( \Sigma _{\pm }-\frac{2\mathcal{P}_{\pm }}{2-\gamma _{\ast }}\right)
^{\prime }=\frac{3}{2}(\gamma _{\ast }-2)(1-\Sigma _{+}^{2}-\Sigma
_{-}^{2})\left( \Sigma _{\pm }-\frac{2\mathcal{P}_{\pm }}{2-\gamma _{\ast }}%
\right) . 
\]%
We can see immediately that $\Sigma _{\pm }-\frac{2\mathcal{P}_{\pm }}{%
2-\gamma _{\ast }}$ is monotone increasing or decreasing according as $%
\gamma _{\ast }>2$ or $\gamma _{\ast }<2$. The equilibrium point $\mathcal{A}%
_{1}$ is a past attractor for $\gamma _{\ast }>2$ and a future attractor for 
$\gamma _{\ast }<2$.

For $\gamma _{\ast }=2$, we can find the exact solution. This is a critical
case in which the equilibrium point disappears. From (\ref{eq:q}) and (\ref%
{eq:omega}), we have $q=2$. The dynamical equations then read 
\begin{eqnarray*}
\Sigma _{\pm }^{\prime } &=&3\mathcal{P}_{\pm }Z, \\
Z^{\prime } &=&-6(\mathcal{P}_{+}\Sigma _{+}+\mathcal{P}_{-}\Sigma _{-})Z,
\end{eqnarray*}%
and give 
\[
\frac{d}{dZ}(\mathcal{P}_{+}\Sigma _{+}+\mathcal{P}_{-}\Sigma _{-})^{2}=-(%
\mathcal{P}_{+}^{2}+\mathcal{P}_{-}^{2}). 
\]%
From this equation, we derive 
\begin{equation}
(\mathcal{P}_{+}\Sigma _{+}+\mathcal{P}_{-}\Sigma _{-})^{2}=-(\mathcal{P}%
_{+}^{2}+\mathcal{P}_{-}^{2})Z+A^{2},
\end{equation}%
where $A^{2}$ is a positive constant. Using this first integral, we arrive
at the exact solution 
\begin{eqnarray}
\Sigma _{\pm } &=&\frac{\mathcal{P}_{\pm }A}{\mathcal{P}_{+}^{2}+\mathcal{P}%
_{-}^{2}}\tanh 3A\tau \pm \mathcal{P}_{\mp }B \\
Z &=&\frac{A^{2}}{\mathcal{P}_{+}^{2}+\mathcal{P}_{-}^{2}}\frac{1}{\cosh
^{2}3A\tau }
\end{eqnarray}%
where $B$ is another integration constant. There is only one independent
parameter because equation (\ref{eq:omega}) serves as a constraint 
\begin{equation}
\frac{A^{2}}{\mathcal{P}_{+}^{2}+\mathcal{P}_{-}^{2}}+(\mathcal{P}_{+}^{2}+%
\mathcal{P}_{-}^{2})B^{2}=1.
\end{equation}

The general solution for $\gamma =2$ corresponds to 
\begin{eqnarray}
a &\propto &e^{\left[ 1+\frac{B}{\sqrt{3}}(\mathcal{P}_{33}-\mathcal{P}_{22})%
\right] \tau }\cosh ^{\frac{3\mathcal{P}_{11}}{\mathcal{P}_{11}^{2}+\mathcal{%
P}_{22}^{2}+\mathcal{P}_{33}^{2}}}3A\tau \\
b &\propto &e^{\left[ 1+\frac{B}{\sqrt{3}}(\mathcal{P}_{11}-\mathcal{P}_{33})%
\right] \tau }\cosh ^{\frac{3\mathcal{P}_{22}}{\mathcal{P}_{11}^{2}+\mathcal{%
P}_{22}^{2}+\mathcal{P}_{33}^{2}}}3A\tau \\
c &\propto &e^{\left[ 1+\frac{B}{\sqrt{3}}(\mathcal{P}_{22}-\mathcal{P}_{11})%
\right] \tau }\cosh ^{\frac{3\mathcal{P}_{33}}{\mathcal{P}_{11}^{2}+\mathcal{%
P}_{22}^{2}+\mathcal{P}_{33}^{2}}}3A\tau .
\end{eqnarray}%
This solution is past and future asymptotic to Kasner vacuum (connecting two
different Kasner points). Kasner exponents are given by 
\begin{eqnarray}
p_{1} &=&\frac{1}{3}+\frac{B}{3\sqrt{3}}(\mathcal{P}_{33}-\mathcal{P}%
_{22})\pm \frac{3A\mathcal{P}_{11}}{\mathcal{P}_{11}^{2}+\mathcal{P}%
_{22}^{2}+\mathcal{P}_{33}^{2}}, \\
p_{2} &=&\frac{1}{3}+\frac{B}{3\sqrt{3}}(\mathcal{P}_{11}-\mathcal{P}%
_{33})\pm \frac{3A\mathcal{P}_{22}}{\mathcal{P}_{11}^{2}+\mathcal{P}%
_{22}^{2}+\mathcal{P}_{33}^{2}}, \\
p_{3} &=&\frac{1}{3}+\frac{B}{3\sqrt{3}}(\mathcal{P}_{22}-\mathcal{P}%
_{11})\pm \frac{3A\mathcal{P}_{33}}{\mathcal{P}_{11}^{2}+\mathcal{P}%
_{22}^{2}+\mathcal{P}_{33}^{2}},
\end{eqnarray}%
where plus sign is for future and minus for past.

\subsection{Invariant set $\Pi (\mathrm{I})$}

On this subset, the governing equations are reduced to 
\begin{eqnarray}
\Sigma ^{\prime } &=&(q-2)\Sigma -\frac{3}{2}(\gamma _{\ast }-2)rZ, \\
\Omega ^{\prime } &=&(2q-3\gamma +2)\Omega , \\
Z^{\prime } &=&\left[ 2q-3\gamma _{\ast }+2+3(\gamma _{\ast }-2)r\Sigma %
\right] Z, \\
q &=&2\Sigma ^{2}+\frac{1}{2}(3\gamma -2)\Omega +\frac{1}{2}(3\gamma _{\ast
}-2)Z, \\
1 &=&\Omega +Z+\Sigma ^{2}.
\end{eqnarray}%
They exhibit significant similarity to the set of equations for Bianchi II.
In fact, the mathematical structure of the equations is the same for both
systems and we can employ the techniques developed for Bianchi II by Collins 
\cite{coll} to analyze our system.\newline

First, by using the constraints we eliminate $q$ and $Z$: 
\begin{eqnarray}
\Omega ^{\prime } &=&3(\gamma _{\ast }-2)\left[ \alpha (1-\Omega )-\Sigma %
\right] \Omega \equiv X,  \label{eq:Omega} \\
\Sigma ^{\prime } &=&\frac{3}{2}(\gamma _{\ast }-2)\left[ \alpha (1-\Omega
)-\Sigma ^{2}+1-\alpha \right] \Sigma -\frac{3}{2}(\gamma _{\ast
}-2)r(1-\Omega -\Sigma ^{2}) \\
&\equiv & Y,
\end{eqnarray}%
For $\alpha >0$, let us define the function $f$ , which was introduced by
Collins \cite{coll}, by 
\begin{equation}
f\equiv \Omega ^{-\frac{3}{2}}\left( 1-\Omega -\Sigma ^{2}\right) ^{-1}.
\end{equation}%
Then, we have 
\begin{equation}
\frac{\partial }{\partial \Omega }(fX)+\frac{\partial }{\partial \Sigma }%
(fY)=\frac{1}{2}\Omega ^{-\frac{3}{2}}\frac{1}{1-\Omega -\Sigma ^{2}}%
(1-\alpha )>0
\end{equation}%
in the interior of $\Pi (\mathrm{I})$. Hence $f$ is a Dulac function for the
system and we conclude that there is no periodic or recurrent orbit in $\Pi (%
\mathrm{I})$ by Dulac's theorem. If $\alpha <0$, we can just take $-f$ as
Dulac's function and are lead to the same conclusion. \newline

Combined with our knowledge of the eigenvalues for the equilibrium points,
we can now draw conclusions about the past asymptotic behaviour of the
system. However, special attention is necessary for the marginal cases with
zero eigenvalues (see ref \cite{skew} for detailed discussion of this
situation in the absence of ultra-critical fluids) and they will be
considered in the next section.

\subsection{Some special cases}

\subsubsection{$r=1:$}

The equilibrium point $\mathcal{A}_{1}$ is identified with the Kasner
equilibrium point $\phi =\theta $. To determine the stability of this point
we need to carry out a second-order analysis. This does not affect the
conclusion that $\Pi (\mathrm{I})$ is the past-asymptotic set and Dulac's
theorem holds for it. Therefore, the past attractor is either $\mathcal{A}%
_{2}$ or $\mathcal{A}_{1}$ for $0<\alpha <1,$ and it is $\mathcal{F}$ for $%
\alpha <0$.

\subsubsection{$\protect\alpha =0:$}

Again, $\Pi (\mathrm{I})$ is past asymptotic set. On $\Pi (\mathrm{I})$ we
can see from equation (\ref{eq:Omega}) that $\Omega $ is monotone increasing
unless $\Sigma =0$. Therefore $\mathcal{F}$ is the past attractor although
linearisation around it produces a zero eigenvalue. It is also clear that $%
\mathcal{A}_{1}$ is a saddle point and $\mathcal{A}_{2}$ is identical to $%
\mathcal{F}$.

\subsubsection{$\protect\alpha =r^{2}:$}

This specialisation ensures $\mathcal{A}_{1}=\mathcal{A}_{2}$ and the
situation is the same as in the general case.

\subsubsection{$\protect\alpha =1:$}

This is the case where $\gamma =2$. The entire $B(\mathrm{I})$ is a set of
equilibrium points with two zero-eigenvalues. There are two sub-cases. If $%
r\leq 1$, $\mathcal{A}_{1}$ is the past attractor while $\mathcal{A}_{2}$ is
unphysical. For $r>1$, $\mathcal{A}_{1}$ is unphysical and $\mathcal{A}_{2}$
is identical to one of the equilibrium points in $B(\mathrm{I})$; some of
them have positive eigenvalues and $\Omega $ is monotone increasing. Thus
all orbits connect two of these equilibrium points.

\subsection{Summary}

For the past asymptotic behaviour of $\overline{B_{A}(\mathrm{I})},$ we have
found the following:

\begin{itemize}
\item For $\alpha \leq 0$, $\mathcal{F}$ is the past attractor.

\item For $0<\alpha <\mathrm{min}\{r^2 , 1\}$, $\mathcal{A}_2$ is the past
attractor.

\item For $\alpha \geq r^2 ,r\leq 1$, $\mathcal{A}_1$ is the past attractor.

\item For $\alpha >1,r>1$, the Kasner circle is the past asymptotic set.

\item For $\alpha =1, r>1$, a part of Kasner disc is the past asymptotic set.
\end{itemize}

\begin{figure}[h]
\begin{center}
\includegraphics[scale=0.3]{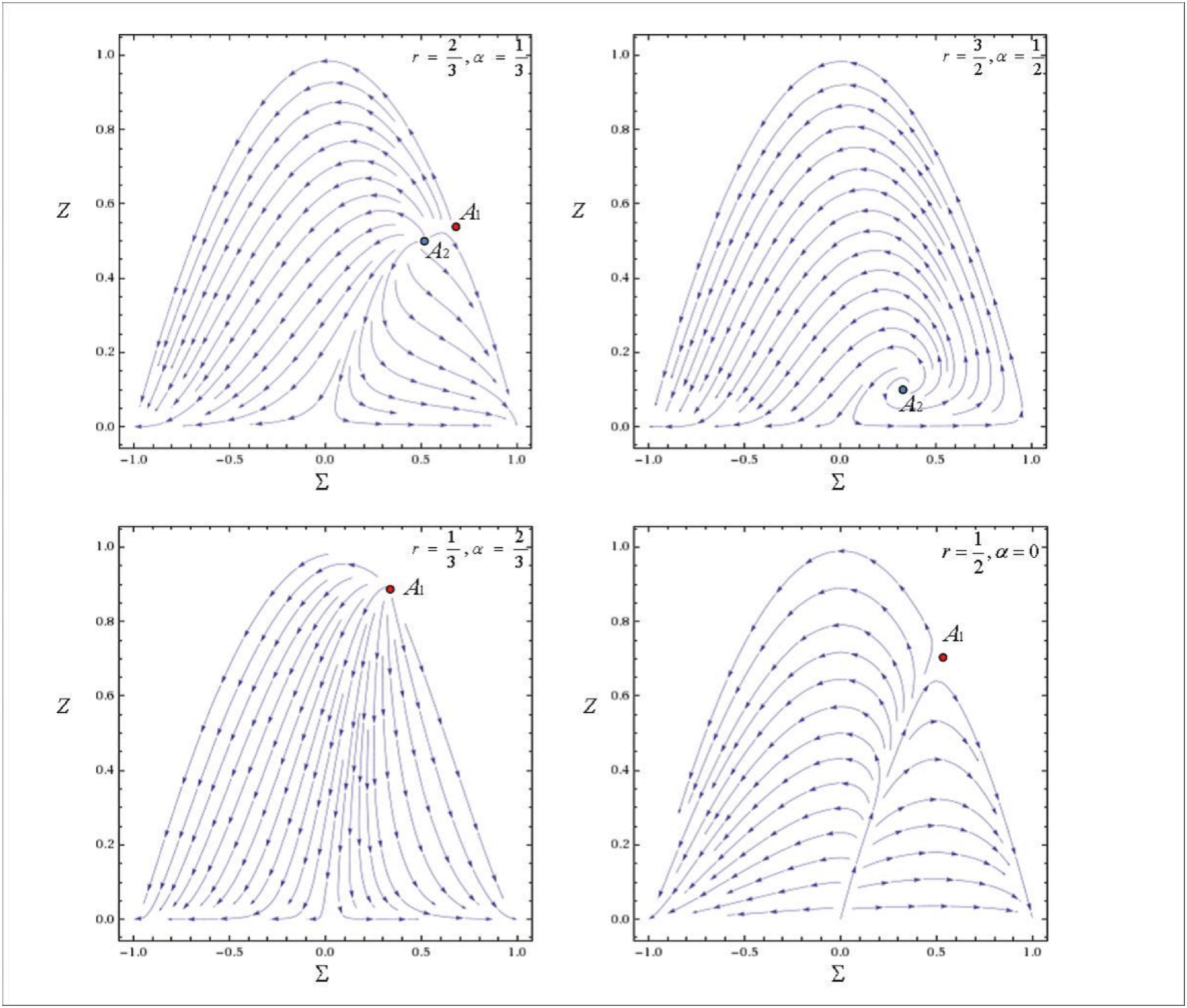}
\end{center}
\caption{Phase portraits for different values of $r$ and $\protect\alpha $.
The Friedmann solution $\mathcal{F}$ is at $(0,0)$ and the Kasner
equilibrium points are located on $(\pm 1,0)$. The critical points $\mathcal{%
A}_{1}$ and $\mathcal{A}_{2}$ are also indicated in the figures.}
\label{fig:2}
\end{figure}

We give a plot in the parameter space of $\alpha $ by $r^{2}$ (Figure \ref%
{fig:1}) indicating the past asymptotic behaviour and showing phase
portraits of the solutions in the invariant set $\Pi (\mathrm{I})$ (Figure %
\ref{fig:2}) for some interesting cases. 
\begin{figure}[h]
\begin{center}
\includegraphics[scale=0.65]{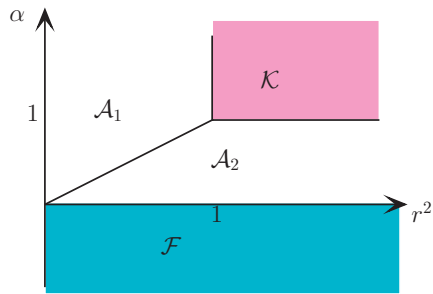}
\end{center}
\caption{The different attractors that are approached near the singularity
for different choices of the parameters $\protect\alpha $ and $r$ when $%
\protect\gamma _{\ast }>2$. The whole parameter space is divided into four
regions. The past attractor is indicated for each of them by the symbol
defined in the text. The isotropic Friedmann attractor is $\mathcal{F}$; the
other attractors are anisotropic.}
\label{fig:1}
\end{figure}

\clearpage

\section{Discussion}

We have studied the behaviour of an anisotropic universe containing
ultra-stiff fluids with isotropic and anisotropic pressures. We find that
the addition of anisotropic ultra-stiff pressures with principal pressures
that can exceed $\rho $ completely changes the results obtained when only an
ultra-stiff perfect fluid with isotropic pressure ($p>\rho $) is present.
Most notably, the isotropic Friedmann universe is no longer the stable
early-time attractor solution as the initial singularity is approached and
the effects of the anisotropic pressures lead to an anisotropic Kasner-like
expansion near the singularity. We would expect that more general Bianchi
type universes with ultra-stiff anisotropic pressures would lead to further
types of anisotropic attractor but the Friedmann singularity would remain
unstable. Any attempt to model the evolution of physical quantities through
a singular or non-singular bounce in a cyclic cosmology will need to be
re-evaluated in the light of these results.

\textbf{Acknowledement: }K. Yamamoto acknowledges support from  the
Cambridge International Scholarship Scheme.

\end{document}